\documentclass[pra,twocolumn,superscriptaddress,showpacs,%
preprintnumbers,amsmath,amssymb]{revtex4}
\usepackage{amsmath}
\usepackage{graphicx}
\usepackage{dcolumn}

\begin{document}

\title{Electron loss of fast projectiles in the collisions with molecules}

\author{V.I. Matveev}\email{matveev.victor@pomorsu.ru}
\affiliation{Pomor State University, 4 Lomonosov St., 163002 Arkhangelsk, Russia}
\author{D.N. Makarov}
\affiliation{Pomor State University, 4 Lomonosov St., 163002 Arkhangelsk, Russia}
\author{Kh.Yu. Rakhimov}
\affiliation{Department of Physics, National University of Uzbekistan, 100174 Tashkent, Uzbekistan}

\date{ \today }

\begin{abstract}
The single and multiple electron loss of fast highly charged
projectiles in the collisions with neutral molecules are studied
within the framework of a nonperturbative approach. The cross
sections for single, double, and triple electron losses are
calculated for the collision system  $Fe^{q+}\to N_2$ ($q$=24,
25, 26) at the collision energies  10, 100, and 1000 MeV/u. The
effects caused by the collision multiplicity and the orientation
of the axis of target molecule are treated.
It is shown that collision multiplicity effect leads to considerable differences for
the cases of perpendicular and parallel orientations of the molecular axes with respect
 to the direction of the projectile motion, while for chaotic orientation such effect is negligible.
\end{abstract}


\maketitle

\section{Introduction}

Stripping, or electron loss by heavy projectiles in the collisions
with atoms has been subject of extensive studies during last
decade.
Various approach for the theoretical study of the projectile electron loss cross sections in the collisions with neutral targets are presented in the Ref.~\cite{Voitkiv1} and in the monograph by A.~Voitkiv and J.~Ullrich~\cite{Voitkiv2} where detailed discussion of the experimental results can be found.
 In theoretical study of such processes it is important the
treatment of the effects caused by strong field of the projectile
and accounting the transitions in the target state. Such a
treatment requires using nonperturbative methods  for the
calculation of stripping cross sections~\cite{DuBois,Olson}. Recently such an approach
was developed on the basis of sudden perturbation approximation
and used for the study of electron losses by fast highly charged
projectiles with neutral gas atoms~\cite{Matveev05,Matveev07}.

Despite the fact that considerable progress is made in theoretical
study of electron loss by fast projectiles, most of the treatments
are restricted by considering atomic targets.

However, study of the electron loss by fast highly charged
projectile in the collisions with molecular targets is of
fundamental and practical importance because of the number of new
effects that cannot be observed in the case of atomic targets.

One of such effects is the strong dependence of the ionization
 (both for projectile and target) cross section, on the orientation
of the molecular axis with respect to the direction of the collision
velocity. Earlier the existence of such an effect was mentioned in
the Refs.~\cite{Siegmann01,Kaliman,Siegmann02}, where the
ionization of molecular targets in their collisions fast highly
charged ions is studied.

Another important effect that appears in the collision of fast
highly charged projectiles with  molecules is so-called
collision multiplicity effect~\cite{Matveev08}. It implies that
after the collision with the first atom of target before the
relaxation  the projectile collides with the
second atom being in the excited state. Considerable contributions
by multiplicity effects to the ionization cross section and energy
losses of fast highly charged ions with diatomic
molecules~\cite{Matveev08} and nanoparticles~\cite{Matveev09} was
found in recently.

In particular, it was shown in the
Refs.~\cite{Matveev08,Matveev09} that the orientation and
multiplicity effects will occur when the time interval between two
subsequent collisions is less  than the relaxation
time. Furthermore, it is clear that the multiple collision of the projectile with target
molecule atoms occurs when the direction of the projectile motion is close to the orientation of target molecule axis.
Qualitatively, the role of the orientation effects can be
understood for diatomic molecule as follows.
Therefore in the study of fast collisions of highly charged projectiles with diatomic molecules
one should take into
account  the two-step processes which include collisions of the
projectile with the first atom of the molecule and cannot relax
into the ground state before the collision with the second atom of
the target.
We note that in the present paper we consider the case when the time between two subsequent collisions is much shorter than the characteristic period of the projectile electrons. Therefore the processes we deal with can be interpreted as  the variation of the "atomic-double-slit" processes in which projectile electrons interacts with both molecular centers in a coherent way (see, e.g., \cite{Voitkiv3}).
It is important to note that the collision multiplicity effects
contribute only to the projectile ionization/excitation cross
section without causing  any changes in the transitions of target
states. Indeed, in subsequent collisions with each atom of the
target molecule the ionization or excitation occurs in different
atoms. It is clear that the above arguments are true for polyatomic
molecules, too.

In this paper we develop a nonperturbative approach which is based
on the use of sudden perturbation approximation and utilizes it for the
calculations of the electron loss cross sections of fast highly
charged projectiles in the collisions with polyatomic molecules.
The method allows to take into account all the transitions both in
target and projectile electronic states. Moreover, it is
possible to achieve considerable simplifications of the
expressions for the electron loss cross sections in the cases of  high
enough projectile charges and multi-electron targets .

The calculations of the electron loss cross sections and their
dependence on the orientation of the molecular axis are presented for
single-, double-, and triple- stripping of the fast highly charged
iron ions. It is also shown that the collision multiplicity
effect leads to considerable difference between the stripping
cross sections for the cases of parallel and perpendicular
orientation of the molecular axis with respect to the direction of projectile
motion. However, for chaotic orientation of the axis of target
molecule  such effect is negligible.
Finally, it should be noted that for the calculation of the multiple stripping cross section of the fast highly charged projectiles in their collisions with multi-electron targets one should use non-perturbative methods, since the Born approximation is not applicable for these processes. Our approach also can be considered as a nonperturbative one. However, our results concerning on the dependence of the cross sections on the molecule's axis  orientation can be obtained within the perturbative approach, too: at least, for a single-electron ionization.

This paper is organized as follows. In the next section we will
present detailed description of the approach to be utilized,
derivation of the nonperturbative expressions for the stripping
cross section and its applications for the calculations of the
multiple stripping cross sections of fast highly charged
projectiles in the collisions with the nitrogen molecule, $N_2$.
The section III presents discussions of the results, while in the
section IV we give some concluding remarks.

\section{Theoretical background}
The system we are going to treat includes a fast highly charged
projectile colliding with a molecule that consists of multi-electron
(number of electrons, $N_A \gg 1$) atoms. For such collision system the cross
section for excitation or ionization  of the projectile from state
$\left|0\right\rangle$ to $\left|k\right\rangle$
(provided arbitrary transition can occur in the target state)
with high accuracy (error is proportional to $1/N_A$) in the
Glauber approximation can be written as~\cite{Matveev07,Matveev08}
\begin{equation}
\sigma = \int\biggl|\bigl\langle k \bigr|\Bigr. \exp\biggl\{-\frac{i}{v}\int\limits_{-\infty}^{+\infty}U dX\biggr\}\Bigl.\bigl|0\bigr\rangle\biggr|^{2}d^{2}{\bf b}\:.
\label{for2}
\end{equation}
Here ${\bf v}$ is the projectile velocity and the $x-$axis is
directed along the vector ${\bf v}$. If ${\bf b}$ is the impact
parameter, and target position is fixed with one of the nuclei
being at the origin of coordinate system, then the position of the
projectiles nucleus can be given by the vector ${\bf R}=({X, \bf
b})$. Atomic units are used here and in the following.
In Eq.~(\ref{for2}) the potential $U$ describes interaction of the
projectile electrons with the target which is considered as a
lengthy object.  In other words, the cross section in
Eq.~(\ref{for2}) is expressed in terms of the charge density of target electrons
~\cite{Matveev08,Matveev09}).
Formally, Eq.~(\ref{for2}) has the same form as that for the ionization cross section in the "frozen" approximation for the target electrons. However, as it was shown in the Ref.~\cite{Olson}, for the multielectron targets ($N_A\gg 1$) this formula describes (with the error order of $1/N_A$) the transition cross section for the projectile electrons summed over the complete set of final states of the target electrons.
 For a molecule
consisting of multielectron atoms the electronic density is almost the same
as the sum of that for isolated atoms.
Therefore for such case
we can consider the target as consisting of isolated and
non-interacting atoms~\cite{Matveev08,Matveev09}). We
describe the charge density of the each atom in the target
within the Hartree-Fock-Slater model~\cite{Salvat} in which
the spatial charge density can be written as
\begin{equation}
\rho_{m}\left(r\right) = -\frac{Z_{m}}{4\pi r}\sum\limits_{i = 1}^{3}A_{m, i}\alpha_{m, i}^{2}
\ e^{-\alpha_{m, i}r}
\:,\qquad \sum\limits_{i = 1}^{3}A_{m, i} = 1\:,
\label{for4}
\end{equation}
where $Z_{m}$ is the charge of the $m$-th atomic nucleus, $A_{m, i}$ and
$\alpha_{m, i}$ are the tabulated constants that can be found in
the Ref.~\cite{Salvat}. The potential created by molecule at the
point ${\bf r}$ can be written in terms of potentials created by
each atom of the molecule:
\begin{equation}
\varphi ({\bf r})=\sum\limits_{m = 1}^{N}\frac{Z_m}{d_m}\Phi_m(d_m)\ ,
\label{V}
\end{equation}
where $N$ is the number of atoms in target,  $d_m$ is the distance
from the $m$ the nucleus in the target to the observation point,
${\bf r}$ and  $\Phi_m(r)$ is the screening function for the
$m$-th atom. In the Hartree-Fock-Slater model screening function
can be written as:
$$
\Phi_{m}(r)=\sum\limits_{i = 1}^{3}A_{m, i}\exp\left(-\alpha_{m, i}r\right).
$$

Furthermore, let us introduce the following notations: ${\bf
r}_p$ is the coordinate of the projectile electrons with respect
to its nucleus,  ($p=1,2,...,N_P$), $N_P$ is the total number of
projectile electrons,  ${\bf R}_m=(X_m,{\bf b}_m)$ are the
distances between the projectile nucleus and the nucleus of the
$m$-th atom of the target. If  ${\bf b}_m$ is the impact parameter
with respect to $m$-th atom, then $d_m=|{\bf R}_{m} + {\bf r}_p|$.

Therefore the interaction potential between the target and
projectile electrons can be written as
\begin{equation}
U = -\sum\limits_{p = 1}^{N_P}\varphi ({\bf r}_p) =
-\sum_{m=1}^{N}\sum_{p=1}^{N_P}\frac{Z_{m}}{\left|{\bf R}_{m} + {\bf r}_p\right|}\Phi_{m}(\left|{\bf R}_{m} + {\bf r}_p\right|)\:.
\label{for3}
\end{equation}

Taking into account  Eq.~(\ref{for3}) for the "eikonal phase"
involved into Eq.~(\ref{for2}) we have
\begin{eqnarray}
\chi =-\frac{1}{v}\int\limits_{-\infty}^{+\infty}U dX  \ \ \ \ \ \ \ \ \ \ \ \ \ \ \ \ \ \ \ \ \ \ \ \
\nonumber \\
=\sum\limits_{m = 1}^{N}\frac{2Z_{m}}{v}\sum_{p=1}^{N_P}\sum\limits_{i = 1}^{3}A_{m, i}K_{0}\left(\alpha_{m, i}\left|{\bf b}_{m} + {\bf s}_p\right|\right)\:,
\label{for7}
\end{eqnarray}
with $K_{0}\left(z\right)$ being the lowest-order McDonald function
and ${\bf s}_p$ is the projection of the vector ${\bf r}_p$  onto the impact parameter plane.
As it was shown in the Ref.~\cite{Matveev07} Eq.~(\ref{for2}) with
the "eikonal phase" given in the formula~(\ref{for7}) is
applicable for relativistic collisions, too. Thus Eq.~(\ref{for2})
describes the cross section for the transition of the projectile
electrons from the initial state $\left|0\right\rangle$ to a state
$\left|k\right\rangle$, under the assumption that arbitrary transitions can
occur in the states of target electrons. The accuracy of the
formula depends on the number of target electrons, $N_A$ i.e., the
error is of order of the quantity $\sim 1/N_A$ for $N_A\gg 1$.

In the following we will consider the highly charged projectiles
whose "visible" (effective) charge, $Z_P$ is much larger than unity (for
instance, for the $Fe^{10+}$ projectile $Z_P=10$ while nucleus
charge is $Z=26$). Then the characteristic size of the projectile
is much less than that of an atom of the target molecule.
Therefore we can assume that the mean field created by atom acts
uniformly to the projectile electrons that corresponds to
expansion of the "eikonal phase" in Eq.~(\ref{for7}) in terms of a
small parameter, $s_p/b$. This allows us to rewrite
Eq.~(\ref{for2}) for orthogonal $\left|0\right\rangle$ and
$\left|k\right\rangle$ in  the following form
\begin{eqnarray}
\sigma = \int\biggl|\left\langle k\right|\exp\biggl(-i\sum\limits_{m = 1}^{N}{\bf q}_m\sum_{p=1}^{N_P}{\bf r}_p\biggr)\left|0\right\rangle\biggr|^{2}d^{2}{\bf b}\;,
\label{sigm}
\end{eqnarray}
where
\begin{equation}
{\bf q}_m=\frac{2Z_{m}}{v}\sum\limits_{i=1}^{3}
\alpha_{m, i}A_{m, i}K_{1}(\alpha_{m, i}b_m)\frac{{\bf b}_m}{b_m}\;,
\label{q_m}
\end{equation}
has the meaning of the momentum transfer to each projectile
electron due to the collisions with $m$-th  atom of the target,
$K_1(z)$ is the first-order McDonald function.

Furthermore, we note that Eq.~(\ref{sigm}) can be  applied for the
collisions  of highly charged ($Z_P\gg 1$) projectiles with the
neutral molecules consisting of multielectron atoms. This requires
fulfilling by $Z_t$ the condition $Z_t\gg 1$, with  $t=1,2,...,N$ being
the number of atoms in target, i.e., $Z_1$ is the nucleus charge of the  first atom of the
target, $Z_N$ is the nucleus charge of the $N$-th atom {\it etc.}

Since the expression for the
cross section is derived within  the Glauber approximation, the
energy of projectile, $E$ should be much larger than that of
projectile-target interaction, $U$, i.e., $E\gg U$ and $kL\gg 1$,
where $k$ is the projectile momentum and $L$ is the interaction
radius for the potential $U$. In the case of neutral molecular
target whose size is much larger than that of projectile, as the
quantity $L$ one can take characteristic size of target. It is
clear that  this condition is fulfilled for fast highly charged
projectile. In addition, to make use our approach we have to
 assume that the sudden perturbation approximation is valid, i.e., for each target
atom (interaction) collision time, $\tau_c \sim L/v$ between the
projectile and target is much shorter than the period of most
fastest (inner) electron, $\tau_e$, to be ionized:
$$
\tau_c \ll \tau_e.
$$

Fulfilling of this condition means that target electrons cannot
change their position during the collision time. In this case
target electrons can considered as in fixed positions during the
collision~\cite{LandauIII}. For relativistic projectiles the
above condition can be written as $\sqrt {1-v^2/c^2}\; L/v \ll\tau_e$. The quantity $\tau_e$ can be estimated for each
fixed collision system. For multielectron, collision system,
with  most of the electrons to be ionized being in outer
shells we have $\tau_e \sim1$.
%
%

As it was mentioned above in order to make use Eq.~(\ref{sigm})
for the calculation of the projectile stripping cross sections in
the collision with neutral molecular targets the following
condition should be fulfilled
$$
Z_P=Z-N_P \gg 1,
$$
where $N_P$ is the number of projectile electrons before the
collision, $Z$ is the charge of the projectile nucleus. This means
that (unscreened) charge of projectile should be large enough to
consider the projectile as highly charged particle. Important
point in the calculations of projectile's multiple stripping cross
section in the collisions with polyatomic molecules is the effects
of collision multiplicity and effects of molecular axis
orientations.  To include these effects into consideration we need
to assume that two or more atoms in the target are on the same
line which is parallel to the vector ${\bf v}$. To demonstrate
collision multiplicity and molecular axis orientation effects we
consider simplest target, diatomic molecule.  We will use notation
${\bf L}$ to denote the orientation vector is of the molecular
axis and assume that it is directed along the line connecting
nuclei of two atoms in the molecule. Then the cross section
presented by formula~(\ref{sigm}) is the function of vector ${\bf
L}$ i.e., $\sigma=\sigma({\bf L})=\sigma(\theta,\phi)$.
Furthermore, we represent vector ${\bf L}$ in terms of spherical
coordinates with the angles $\phi$, $\theta$ and assume the
molecular axis to be directed along the vector ${\bf v}$.

In the following we will be interested in the stripping cross
section averaged over the angle $\phi$:
\begin{equation}
 \sigma(\theta)=\frac{1}{2\pi}\int \sigma(\theta,\phi) d\phi \ .
 \label{st}
\end{equation}

Angle $\theta$ will be called the orientation angle of the
molecular axis. Then the multiplicity effect can be characterized
in terms of $\theta-$dependent relative correction, $\delta$
defined as $\delta=(\sigma(\theta)-\sigma_{\bot})/\sigma_{\bot}$,
where $\sigma_{\bot}$ is the stripping cross section at
$\theta=\pi/2$. It is easy to see that
 $\sigma_{\bot}$ described the collision when target moleculae
 axis is perpendicular to the projectile direction.

\section{Results and discussion}

Let us now apply Eq.~(\ref{sigm}) to the collision system consisting of fast $Fe^{q+}$ projectile and $N_2$ molecule.
For simplicity we consider the cases when $q=24, 25, 26$.
Let us start from the most simplest case: stripping of hydrogen-like
projectile in the collision with diatomic molecule for which the
cross section can be written as
\begin{equation} \sigma(\theta,\phi)=\int d^2 b\int
d{\bf k} \mid\langle{\bf k}\mid \exp\{-i({\bf q}_1+{\bf q}_2){\bf
r}\}\mid0\rangle\mid^{2}\,. \label{H}
\end{equation}
Here ${\bf r}$ is the
coordinate of the projectile electron with respect to the
projectile nucleus,
 ${\bf k}$ is the momentum of the projectile electron lost ${\bf
q}_j$ is the momentum transfer that can be written as ${\bf
q}_j=\frac{2Z_A}{ v}\sum_{i=1}^{3}\alpha_{i}A_{i}K_{1}(\alpha_{i} b_{j})
\frac{{\bf b}_j}{b_j}$, where  ${\bf b}_j$ is the impact parameter with respect
to $j$th nucleus of the target ($j=1,2$).

Using Eqs.~(\ref{st}) and~(\ref{H}) we have calculated the
stripping cross section $\sigma(\theta)$, of the hydrogen-like
projectile   $Fe^{25+}$ in the collision with nitrogen molecule
 $N_2$, for different orientation angles and collision energies.
A quantity we are interested to analyze is the correction to the
projectile stripping cross section due to the collision
multiplicity which is given as
$\delta(\theta)=(\sigma(\theta)-\sigma_{\bot})/\sigma_{\bot}$. In
Fig.~1 the results of calculation of such quantity for collision
system $Fe^{25+}-N_2$ are presented. Continuous line in this
figure represents $\delta(\theta)$ for the collision energy 10~MeV/u, while long-dashed and short-dashed lines are the results
for the energies 100~MeV/u and 1000~MeV/u, respectively. The
angle $\theta$ is given in radians.

\begin{figure}[!bpht]
\centerline{\includegraphics[width=.8\linewidth]{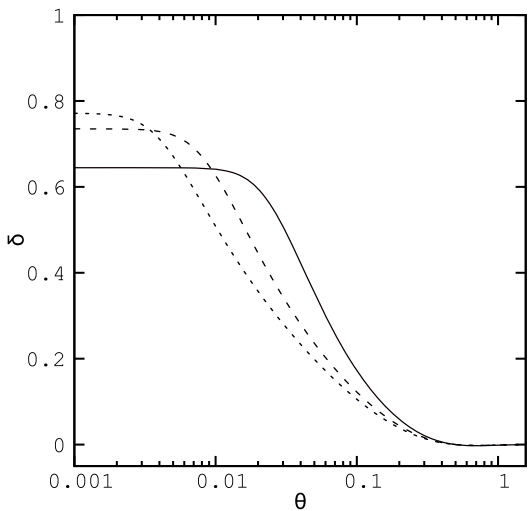}}
\caption{The dependence of the quantity
$\delta(\theta)=(\sigma(\theta)-\sigma_{\bot})/\sigma_{\bot}$ on
the orientation o target molecule axis $\theta$ for the collisions
system $Fe^{25+}\to N_2$ with $\sigma_{\bot}=\sigma (\theta
=\pi/2)$. Continuous line is the result for the collision energy
10 MeV/u; long-dashed line is the result for the collision energy
100 MeV/u and short-dashed line is $\delta (\theta)$ for the energy
1000 MeV/u. The values of $\theta$ are given in radians.}
\label{fig:1}
\end{figure}

Now let us consider stripping of helium-like projectile
$Fe^{25+}-N_2$, in the collision with $N_2$ molecule.  In this
case according to Eq.~(\ref{sigm}) the cross section for the
transition of projectile from state $|0, 0\rangle$ to $|n_1, n_2\rangle$ can
be written as
\begin{equation}\sigma(\theta,\phi)=\int d^2 b
\mid\langle{n_1},{n_2}\mid e^{-i{({\bf q}_1+{\bf q}_2)}({{\bf
r}_{1}}+{{\bf r}}_{2})}\mid0,0\rangle\mid^{2}\,, \label{He}
\end{equation}
with  ${\bf r}_{1}$ and ${\bf r}_{2}$ being the coordinates of the
projectile electrons with respect to projectile nucleus.

Following the Refs.~\cite{Matveev09PR,Matveev98R}) in our
calculations (final and initial) two-electron states of the
projectile  are described as a symmetric products of one-electron
hydrogen-like wave functions with effective charges equal to degree of ionization. Single $\sigma^{1+}(\theta)$, and
double $\sigma^{2+}(\theta)$, electron loss cross sections of can
be calculated using Eq.~(\ref{He}).
Here the cross section for double stripping $\sigma^{2+}(\theta)$, corresponds to
transition of both electron into the continuum state, while the single projectile
ionization cross section $\sigma^{1+}(\theta)$,  describes transition of one of the
electrons into  the continuum, by exciting  another one into a state of the discrete spectrum.
Correspondingly, the quantity
$\delta$ can be estimated for both cases. In Figs.~2a, 2b, and 2c
we plotted  $\delta$ as a function of orientation $\delta$ for the
collision energies 10, 100, and 1000~MeV/u, respectively.
Continuous line in this figure is the result for single electron
loss,
$\delta=(\sigma^{1+}(\theta)-\sigma^{1+}_{\bot})/\sigma^{1+}_{\bot}$,
while dashed line describes calculation of $\delta$ for double
ionization,
i.e., $\delta=(\sigma^{2+}(\theta)-\sigma^{2+}_{\bot})/\sigma^{2+}_{\bot}$.

\begin{figure}[!bpht]
\centerline{\includegraphics[width=0.8\linewidth]{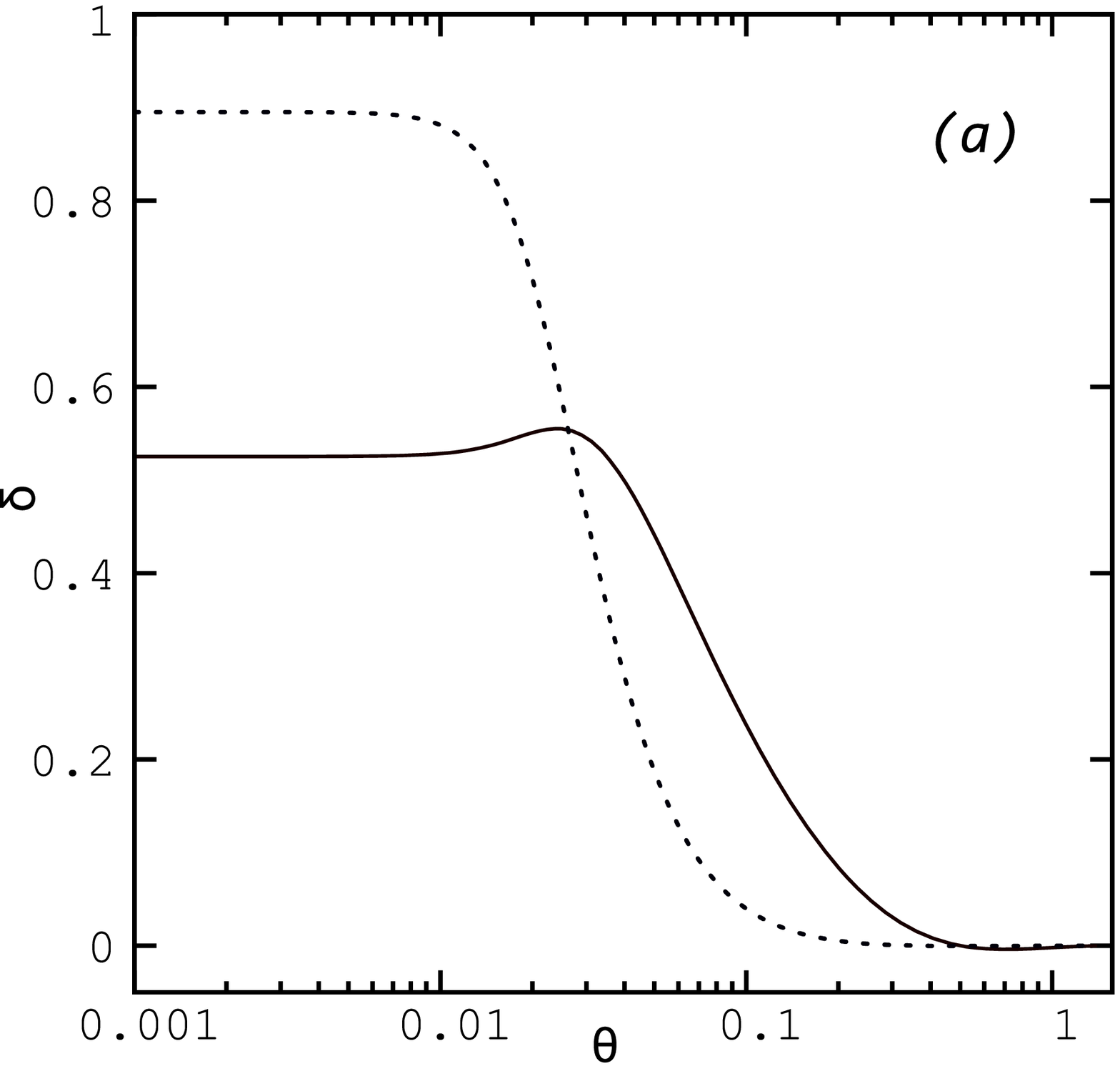}}
\centerline{\includegraphics[width=0.8\linewidth]{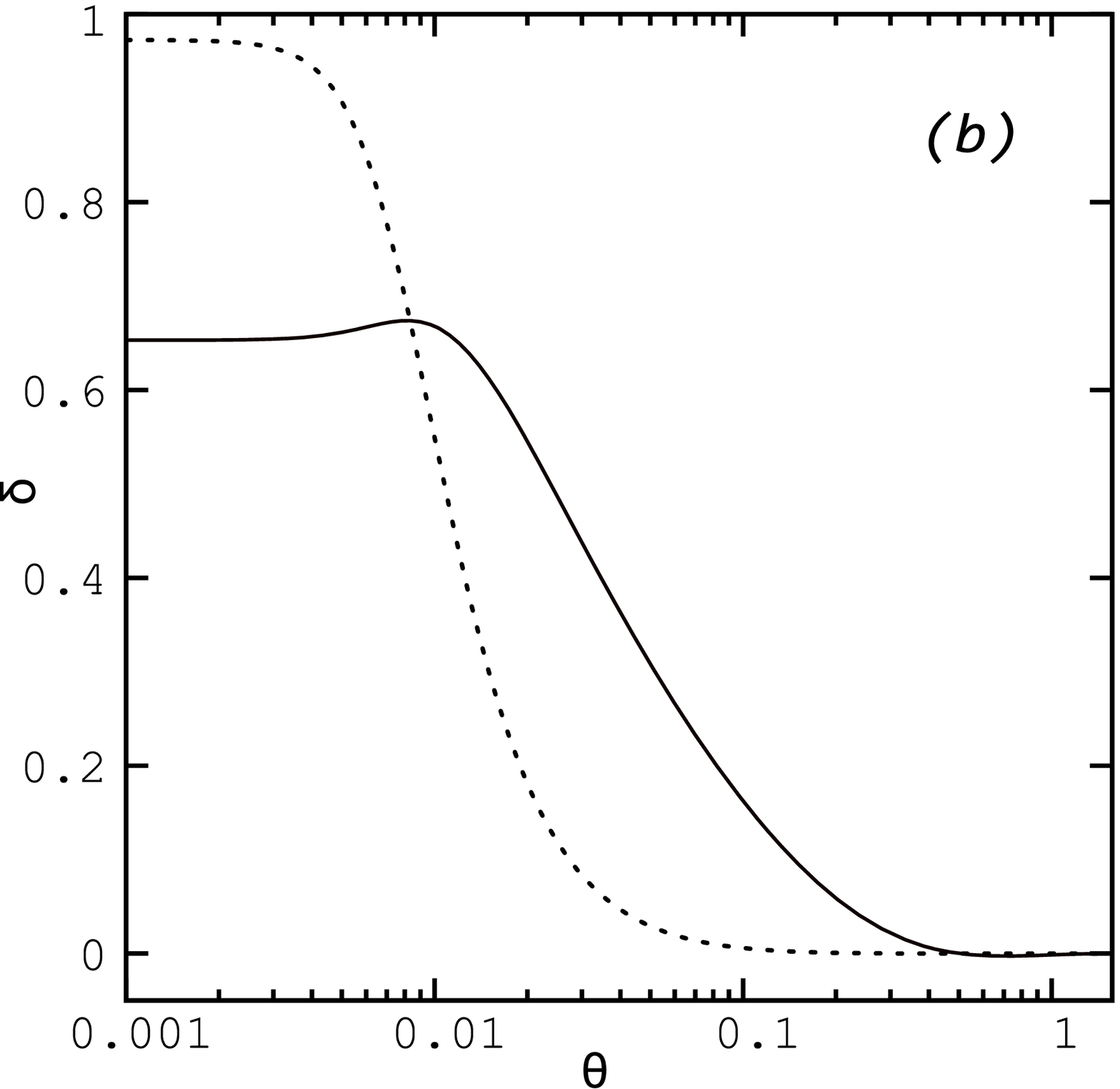}}
\centerline{\includegraphics[width=0.8\linewidth]{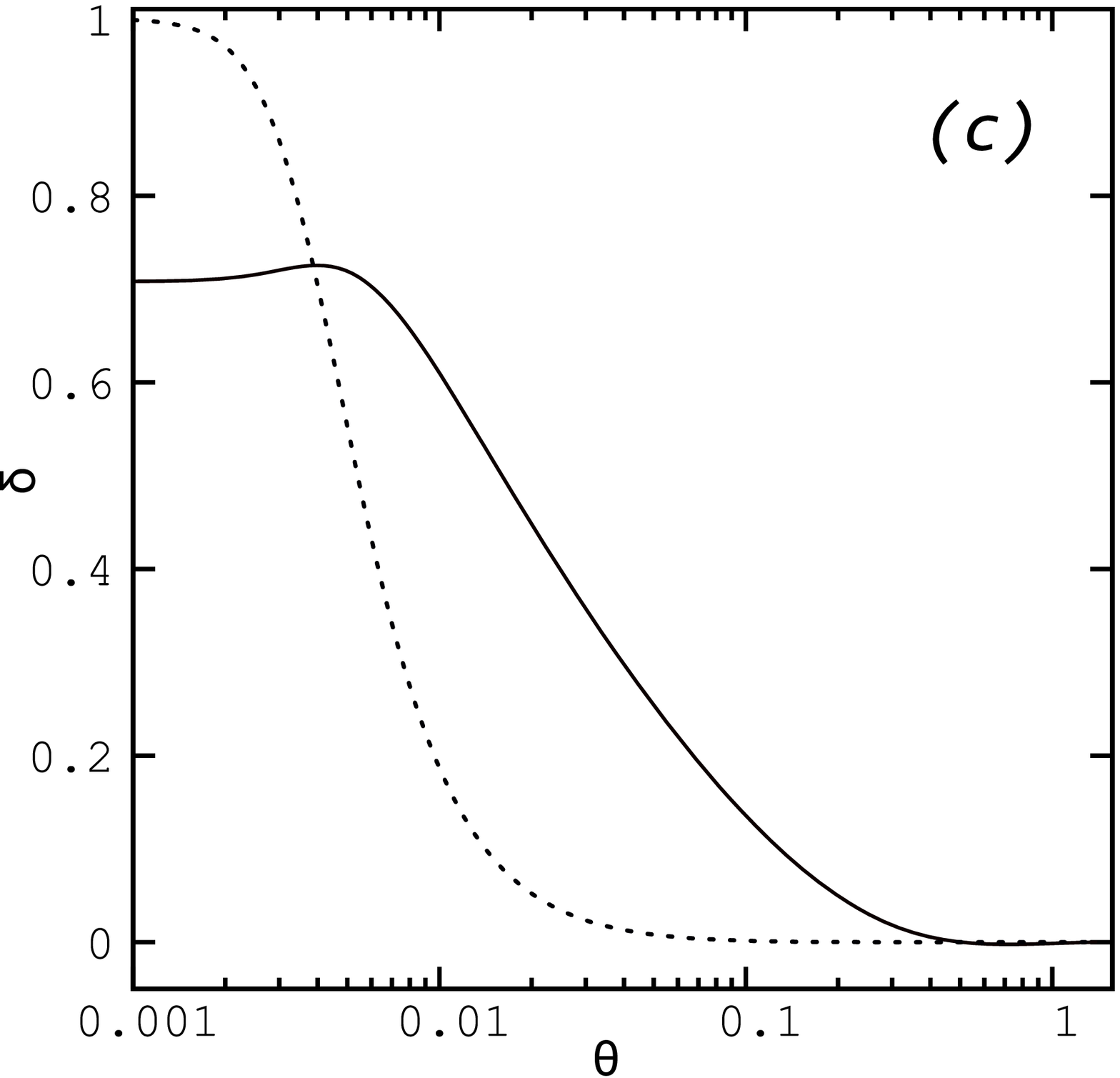}}
\caption{
The dependence of the quantity
$\delta(\theta)=(\sigma(\theta)-\sigma_{\bot})/\sigma_{\bot}$ on
the orientation o target molecule axis $\theta$ for the collision
system $Fe^{24+}\to N_2$ for the energies 10 (Fig.~2a), 100
(Fig.~2b), and 1000 MeV/u (Fig.~2c). Continuous line is the result
for single electron loss, dashed line is the result for double
electron loss by projectile.}
\label{fig:2}
\end{figure}

Finally, we utilized Eq.~(\ref{sigm}) for the calculation of
multiple stripping of three-electron (Lithium-like) projectile,
Fe$^{23+}$ in the collision with $N_2$ molecule. In this case the
loss cross section describing transition from state $|0,0,0\rangle$ to
$|n_1,n_2,n_3\rangle$ can be written as
\begin{equation}\sigma=\int d^2 b
\mid\langle n_1,n_2,n_3\mid e^{-i{({\bf q}_1+{\bf q}_2)}({{\bf
r}_{1}}+{{\bf r}}_{2}+{{\bf r}}_{3})}\mid0,0,0\rangle\mid^{2}\,,
\label{Li}
\end{equation}
where  ${\bf r}_1$, ${\bf r}_2$, and ${\bf r}_3$ are coordinates of projectile electrons.

As in the case of helium-like projectile, the wave functions of
three electron state are taken as the symmetric product of
one-electron (hydrogen-like) wave functions with effective charges equal to degree of ionization discussed in the Refs.~\cite{Matveev09PR,Matveev98R}).
Single, double, and triple electron loss cross sections of
Fe$^{23+}$ projectile can be calculated using Eq.~(\ref{Li}).
Single electron loss  implies ionization of one electron, while
other two electrons can be excited into any state of the discrete
spectrum. Similarly, in the case if double electron loss two
electrons are lost, while other electron excited into any bound
state. As in the cases of single and double electron losses, we
have calculated the quantity, $\delta$ for the cases of single-,
double-, and triple-electron losses. Figs.~3a, 3b, and 3c present
the results of calculation of $\delta(\theta)$ for the collision
energies 10, 100, and 1000 MeV/u, respectively. Continuous line in
this figure is the result for single electron loss,
$\delta=(\sigma^{1+}(\theta)-\sigma^{1+}_{\bot})/\sigma^{1+}_{\bot}$,
long dashed line describes calculation of $\delta$ for double
ionization,
$\delta=(\sigma^{2+}(\theta)-\sigma^{2+}_{\bot})/\sigma^{2+}_{\bot}$
and short-dashed line is the plot of
$\delta=(\sigma^{3+}(\theta)-\sigma^{3+}_{\bot})/\sigma^{3+}_{\bot}$.

\begin{figure}[!bpht]
\centerline{\includegraphics[width=0.8\linewidth]{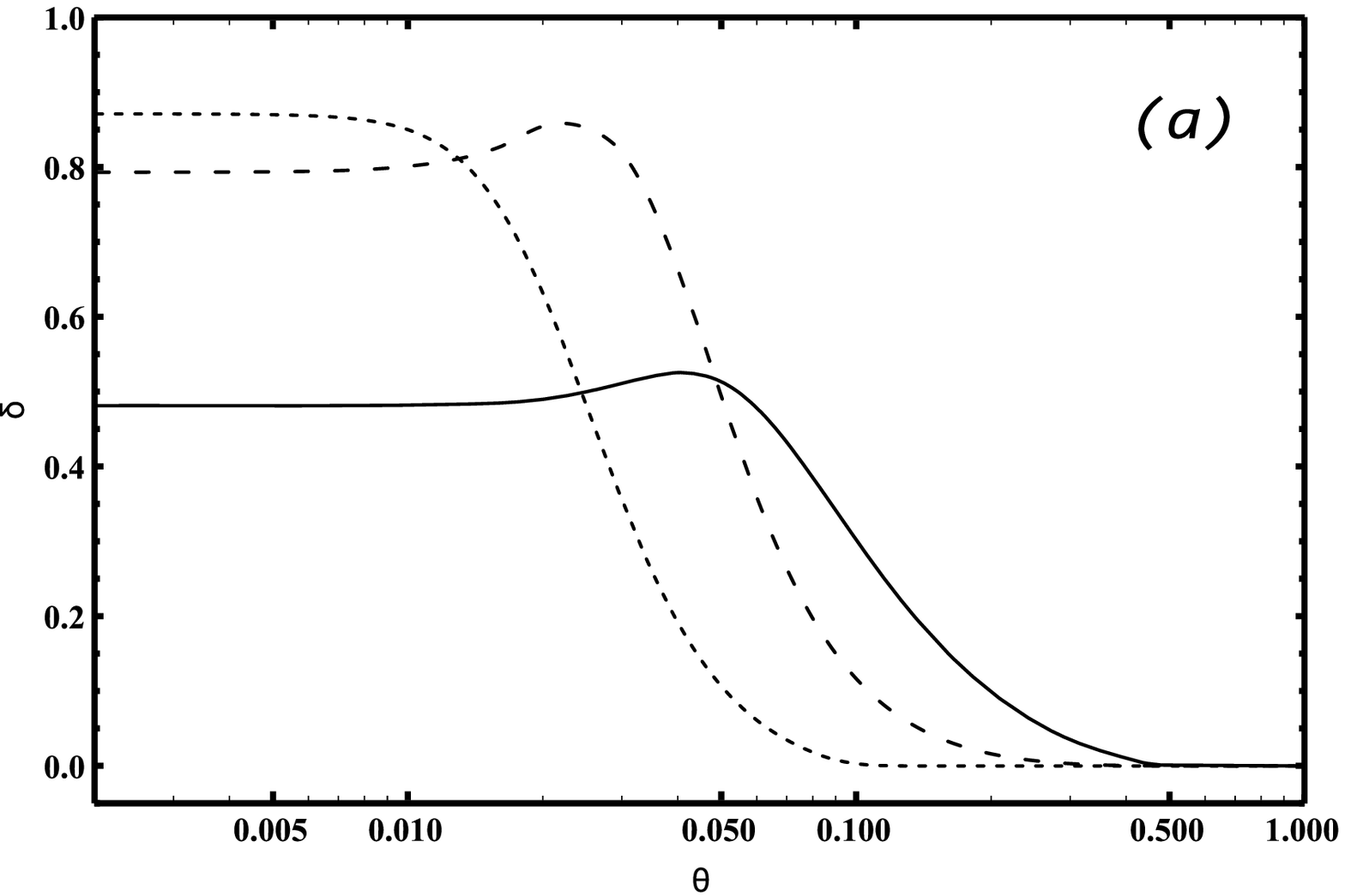}}
\centerline{\includegraphics[width=0.8\linewidth]{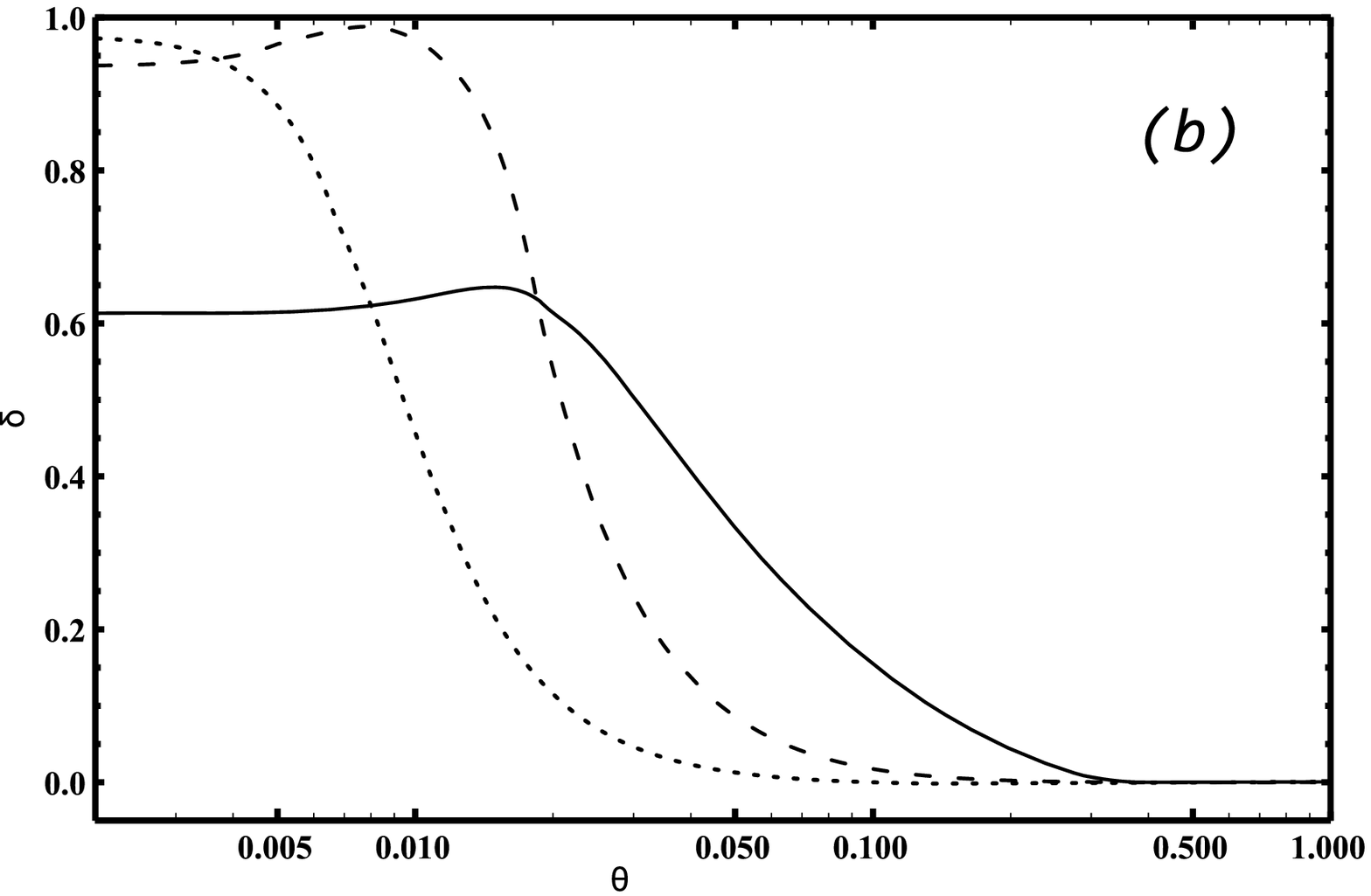}}
\centerline{\includegraphics[width=0.8\linewidth]{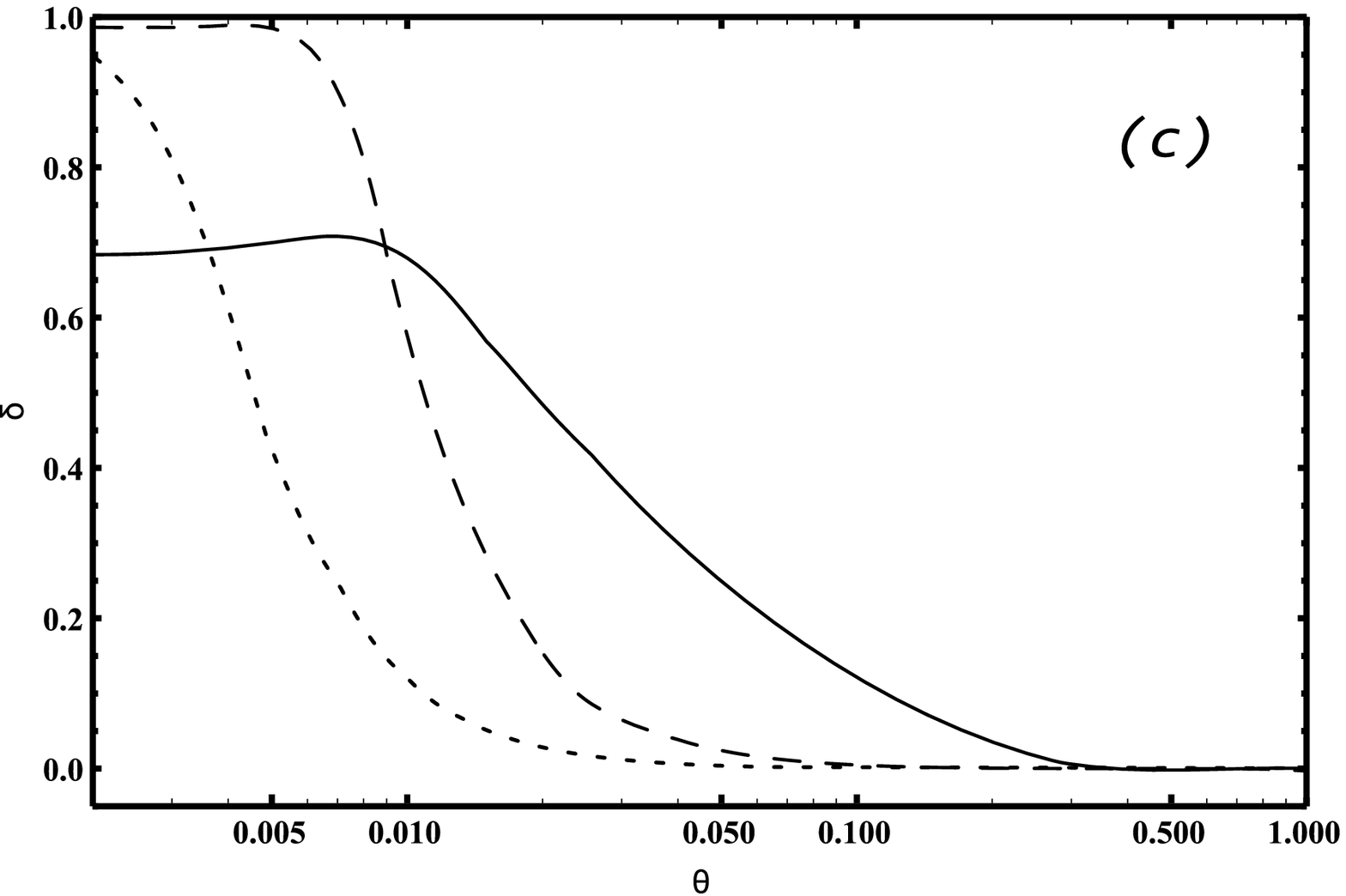}}
\caption{The dependence of the quantity
$\delta(\theta)=(\sigma(\theta)-\sigma_{\bot})/\sigma_{\bot}$ on
the orientation o target molecule axis $\theta$ for the collision
system $Fe^{23+}\to N_2$ for the energies 10 (Fig.~3a), 100
(Fig.~3b), and 1000 MeV/u (Fig.~3c). Continuous line is the result
for single electron loss, long-dashed line is the result for
double electron loss and short-dashed line for triple electron
loss by projectile.}
\label{fig:3}
\end{figure}

As is seen from the Figs.~1-3, for all collision energies stripping
cross sections considerably depend on the orientation angle
$\theta$, the difference between "perpendicular" and "parallel"
orientations is about 50\%-100\%. It is easy to estimate from the
geometrical analysis the value of the angle $\theta$ at which
multiplicity effect becomes essential.  Since in all cases
projectile collides with the neutral atoms of the target molecule
whose sizes $\sim 1$, for the internuclear distance of target
atoms denoted by $L$, the orientation can be estimated as  $\theta
\leq 1/L$. Therefore for the nitrogen molecule, for instance, we
have $L=2.07$, that gives the estimate $\theta \leq 0.5$. This can
be seen also from the Figs.~1-3.

Also, it should be noticed that the behavior of the quantity
$\delta$ almost the same for all collisions partners. To check
this we have done (similar to the above) calculations for one-,
two-, and three-electron  $Ni$ and  $Xe$ projectiles colliding with
$N_2$, $O_2$, and $Au_2$ targets. The difference  from the above
treated collision systems we observed for these systems was too
small to present them in this paper.

Finally, since in the experiment the cross section for chaotic
orientation are usually measured, we have done calculations of the
stripping cross section averaged over the angle $\theta$ assuming
uniform distribution on the orientation angle:
\begin{equation}
\overline{\sigma} =\int \sigma({\bf L}) \frac{d\Omega}{4\pi}=\int\limits_{0}^{\pi}\sigma(\theta) \frac{1}{2}\sin\theta\,d\theta \:.
\label{sigmSR}
\nonumber
\end{equation}

As showed the results of such calculations the corrections due to
the collision multiplicity for the case of chaotic orientation is
too small: order of 0.1 percents. In this case the difference
between the cross sections $\overline{\sigma}$ and
$\sigma_{\bot}$ is very small. Thus the collision multiplicity
effect is considerable only for the case of regular (non-chaotic)
orientation of the molecular axis, while for chaotic orientation
it becomes negligible.

\section{Conclusions}
The electron losses of fast highly charged projectiles in the
collisions with neutral molecules  has been studied. Based on the
Glauber approximation a nonperturbative approach is developed to
estimate single and multiple stripping cross sections. Using the
method single, double, and triple electron loss cross sections of
the fast $Fe^{25+}$, $Fe^{24+}$, and $Fe^{23+}$ ions in the
collisions with $N_2$ molecule are calculated. The effect of
collision multiplicity, caused by the collisions of the projectile
with separate atoms of the target is analyzed. It is shown that
multiplicity effect is essential for the case when the orientation
of the target molecule axis is parallel or perpendicular to the
projectile velocity direction, while for chaotic orientation such
effect is negligible.
We note that in all the cases we calculate the total electron loss cross sections, i.e., we perform summation (integration) over the all continuum states of the electron after the ionization. Therefore, in the present paper we didn't discuss the energies and directions of the emitted electrons.
 The above developed method is rather simple
for application and can be used for any (polyatomic) target
molecules and for the projectiles of arbitrary high (including
relativistic) velocities.

\section{Acknowledgement}
We are grateful to Davron Matrasulov for valuable discussions and useful comments.
This work is supported by the grant of Russian Foundation of Basic
Studies  (Ref.~Nr.~08-02-00711-a) and by the grant of the Russian Federation  President (Ref.~Nr.~MK-3592.2011.2).

\end{document}